\documentclass{ws-mpla}
\usepackage{color}
\usepackage{graphicx}
\usepackage{dcolumn}
\usepackage{bm}

\usepackage[sort]{cite}

\newcommand{\be}{\begin{equation}}
\newcommand{\ee}{\end{equation}}
\newcommand{\bea}{\begin{eqnarray}}
\newcommand{\eea}{\end{eqnarray}}

\begin{document}

\markboth{Xiao-Gang He} {A Brief Review on Dark Matter Annihilation Explanation for $e^\pm$ Excesses in Cosmic Ray}


\begin{center}
{\Large {\bf A Brief Review on Dark Matter Annihilation\\ Explanation for $e^\pm$ Excesses in Cosmic Ray}}
\end{center}

\begin{center}
{Xiao-Gang He$^{1,2}$}\\
\vspace*{0.3cm}
$^1$Center for High Energy Physics, Peking University, Beijing\\
$^2$Department of Physics, Center for Theoretical Sciences, and LeCospa Center\\
National Taiwan University, Taipei
\end{center}


\begin{center}
\begin{minipage}{12cm}

\noindent Abstract:\\
Recently data from PAMELA, ATIC, FERMI-LAT and HESS show that there are $e^{\pm}$
excesses in the cosmic ray energy spectrum. PAMELA observed excesses only in $e^+$, but not in anti-proton spectrum.
ATIC, FERMI-LAT and HESS observed excesses in $e^++e^-$ spectrum, but the detailed shapes are different which requires future
experimental observations to pin down the correct data set. Nevertheless a lot of efforts have been made
to explain the observed $e^\pm$ excesses, and also why PAMELA only observed excesses in $e^+$ but not in anti-proton.
In this brief review we discuss one of the most popular mechanisms to explain the data, the dark matter
annihilation. It has long been known that about 23\% of our universe is made of relic dark matter.
If the relic dark matter was thermally produced, the annihilation rate is constrained resulting in the need of a large boost factor to explain the data.
We will discuss in detail how a large boost factor can be obtained by the
Sommerfeld and Briet-Wigner enhancement mechanisms. Some implications for particle physics model buildings will also  be discussed.

\end{minipage}

\end{center}

\setcounter{footnote}{0}
\vskip2truecm

\section{Introduction}

\begin{figure}[t]
\includegraphics[width=0.45\columnwidth
]{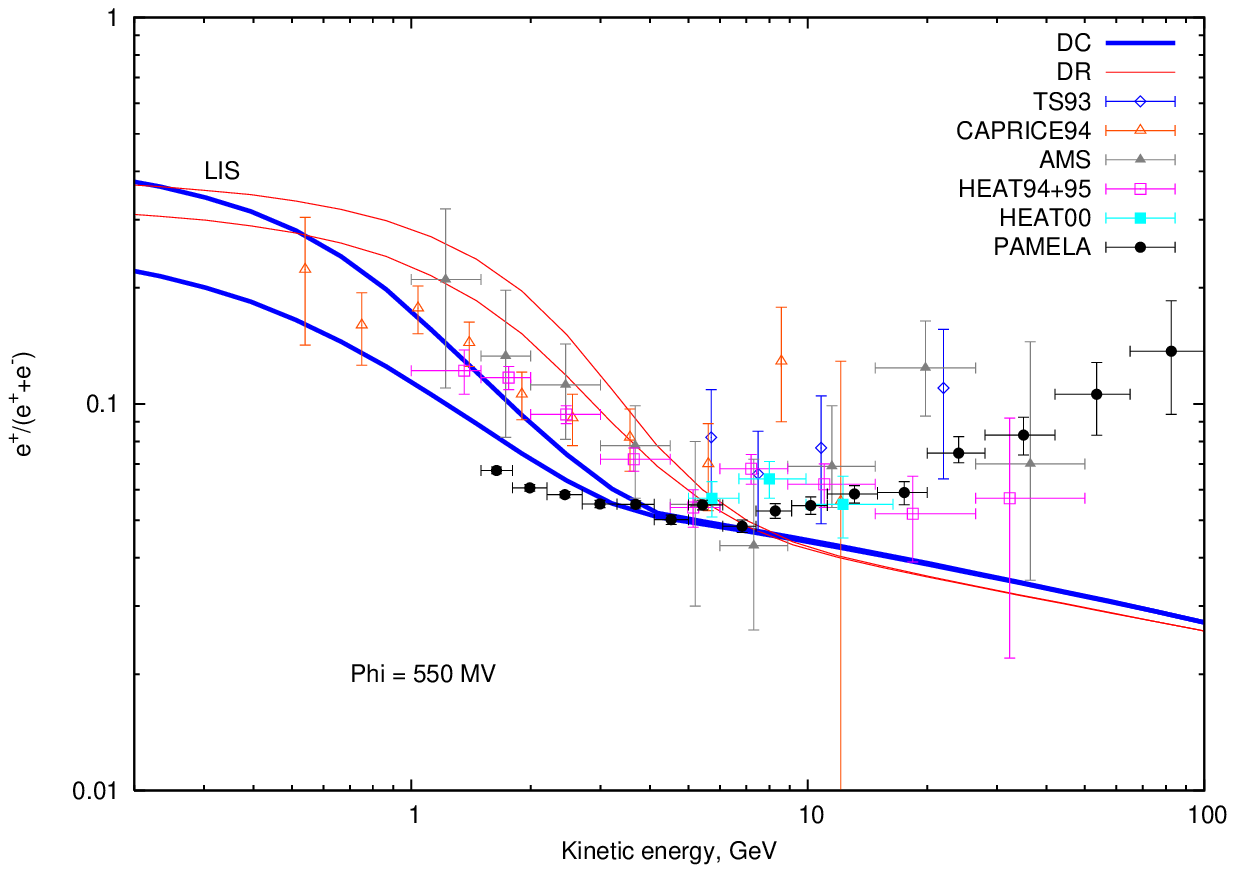}
\includegraphics[width=0.45\columnwidth
]{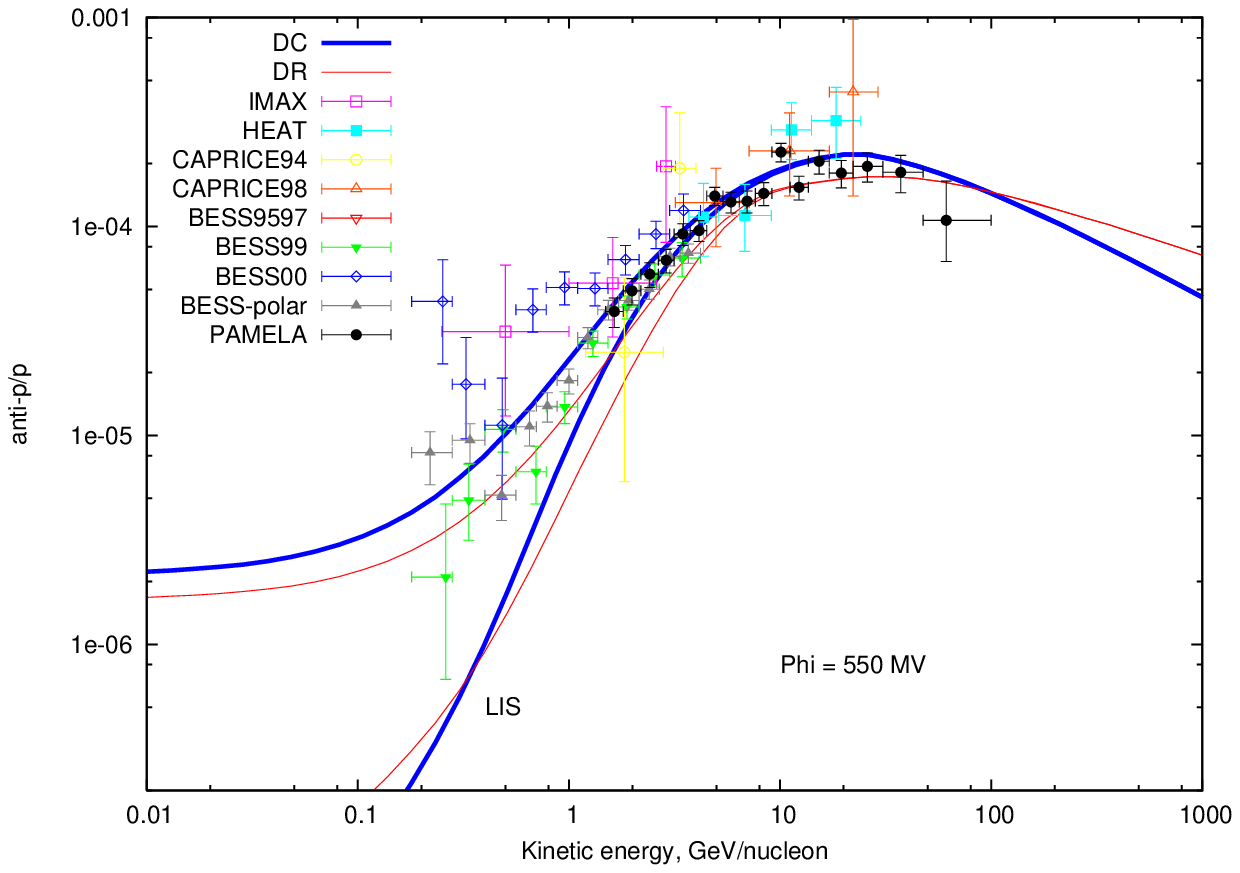}
\caption{
\label{data-pamela}
Observational data and background estimate
on $e^+$ (left) and $\bar p$ (right) energy spectra.
}
\end{figure}

\begin{figure}[t]
\begin{center}\includegraphics[width=0.5\columnwidth
]{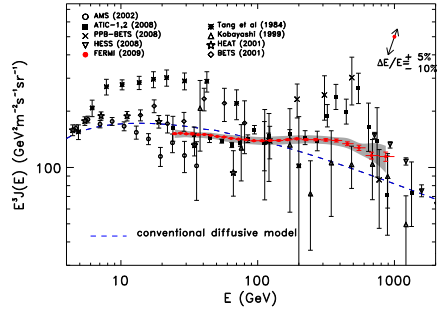}\end{center}
\caption{
\label{data-fermi}
Observational data and background estimate on $e^+ + e^-$ energy spectrum. $J$ is the total flux of $e^+ + e^-$.
}
\end{figure}

Recently several experiments have reported $e^\pm$ excesses in
cosmic ray energy spectrum. Last year the PAMELA collaboration
reported $e^+$ excesses in the cosmic ray energy spectrum from 10 to
100 GeV, but observed no anti-proton excess~\cite{pamela-e,pamela-p}
compared with predictions from cosmic ray
physics~\cite{calprop,other,background,xiao-jun1}. These results are
compatible with the previous HEAT and AMS01 experiments (e.g., Ref.
\cite{heat,ams}) but with higher precision. Shortly after the ATIC
and PPB-BETS balloon experiments have reported excesses in the $e^+
+ e^-$ spectrum between 300 and 800 GeV~\cite{atic,ppb-bets}. The
ATIC data show a sharp falling in the energy spectrum around 600
GeV. Newly published result from FERMI-LAT collaboration also shows
excesses in the $e^+ + e^-$ energy spectrum above the
background~\cite{fermi}. However, the spectrum is softer than that
from ATIC. In addition, the HESS collaboration has inferred a flat
but statistically limited $e^++e^-$ spectrum between 340~GeV and
1~TeV~\cite{hess1} which falls steeply above 1~TeV~\cite{hess2}. The
summary of data are shown in Fig. \ref{data-pamela} and Fig.
\ref{data-fermi} adapted from references \cite{xiao-jun1} and
\cite{fermi}.

Although astrophysics calculations of background $e^\pm$ spectrum in
our galaxy have errors due to model
parameters~\cite{calprop,other,background,xiao-jun1}, within
reasonable ranges it is not possible to eliminate the excesses in
the energy range from 10 GeV to 1 TeV. The $e^\pm$ excesses in
cosmic ray have generated much excitement in particle physics
community because dark matter (DM), which contributes about 23\%
energy density of our universe with properties different than those
of the standard model (SM) particles, can provide a nature
explanation~\cite{strumia,strumia1,model-independent,m-ind1,m-ind2,m-ind3,m-ind4,m-ind5,
m-ind6,m-ind7,m-ind8}\cite{arkani}\cite{leptophilic,lep1,lep2,lep3,bi-he}\cite{susy,susy1,susy2,susy3,susy4,
susy5,susy6,susy7,
susy8,susy9,susy10,susy11,susy12,susy13,susy14,
susy15,susy16,susy17}\cite{nmssmhooper,nmssmhooper1,nmssmlykken}\cite{
kk,kk1,kk2}\cite{light,light1,light2, light3,light4,light5,
light6,light7,light8, light9,light10,light11,
light12,light13,light14, light15,light16,light17,
light18,light19,light20, light21,light22,light23}\cite{annihilation,
annihilation1, annihilation2, annihilation3, annihilation4,
annihilation5, annihilation6, annihilation7, annihilation8,
annihilation9, annihilation10, annihilation11, annihilation12,
annihilation13, annihilation14,
annihilation15,annihilation16}\cite{decay,decay1,decay2,decay3,decay4,
decay5,decay6,decay7,decay8, decay9,decay10,decay11,decay12,
decay13,decay14,decay15,decay16, decay17,decay18,decay19,decay20,
decay21,decay22,decay23,decay24, decay25,decay26,decay27,decay28,
decay29,decay30,decay31,decay32,decay33,decay34}\cite{nonthermal,nonthermal1,nonthermal2}\cite{xiao-jun2,
xiao-jun21,xiao-jun22,xiao-jun23}.



If the data from recent PAMELA, ATIC, FERMI-LAT and HESS are confirmed, one can extract a lot of information about dark matter.
The mass of the annihilating DM serves as the cut off
scale of the $e^\pm$ spectrum,
the lepton spectra must have a cut off energy at the DM mass $m_D$. The FERMI-LAT and HESS data would require that the DM mass to be around 1 to 2 TeV. The DM belongs to
the weakly interacting massive particle (WIMP) category.

To produce  large enough excesses with the annihilation mechanism, it requires modifications of the usual DM
properties. This is because that for the usual DM the annihilation rate producing the $e^\pm$ excess signal
is also related to the annihilation rate producing the cosmological relic DM density. The latter requires that the thermally averaged annihilation rate $<\sigma v>$ to be
$3\times 10^{-26}$ cm$^3$s$^{-1}$. This annihilation rate is too small by a factor of 100 to 1000 to explain the observed excess.
There is the need to boost up the spectrum with a boost factor~\cite{other,background,xiao-jun1} $B$ of 100 to 1000.

Several mechanisms have been proposed to produce a large boost
factor, including the DM
substructures~\cite{xiao-jun2,xiao-jun21,xiao-jun22,xiao-jun23},
non-thermal production of DM
mechanism~\cite{nonthermal,nonthermal1,nonthermal2},
the Sommerfeld
mechanism~\cite{strumia1}\cite{arkani}\cite{nojiri,nojiri1,nojiri2,s-new,sommerfeld-other,sommerfeld-other1},
and the Breit-Wigner
mechanism~\cite{strumia1}\cite{bi-he}\cite{breit,breit1}.
There are also some other proposals~\cite{other-p, other-p1,
other-p2, other-p3}.
Detailed calculations based on the N-body simulation show that the
boost factor from DM substructures is generally less
than~\cite{xiao-jun2,xiao-jun21,xiao-jun22,xiao-jun23}
$\sim 10$.
The reason why non-thermal production of DM in the early universe
can explain $e^\pm$ excesses is because that in this scenario the interaction rate responsible for the excesses are not
directly related to the relic DM density. One basically assumes that the interaction rate is that
required from the $e^\pm$ excess data, and therefore there is no need of a boost factor.
The Sommerfeld and Breit-Wigner mechanisms are more of particle physics answers to the problem which require the existence of new particles.
In order for the Sommerfeld mechanism to be in effective, the new particle needs to be light to allow long range interaction between DM.
The Breit-Wigner mechanism works if the annihilation of DM is through s-channel
and the new particle to have a mass twice of the DM mass. This is a resonant effect.

If the DM is not stable and decay predominately into leptons on time
scale longer than the age of the universe, DM decay can also provide
another alternative explanation to the
data~\cite{xiao-jun1}\cite{decay,decay1,decay2,decay3,decay4,
decay5,decay6,decay7,decay8, decay9,decay10,decay11,decay12,
decay13,decay14,decay15,decay16, decay17,decay18,decay19,decay20,
decay21,decay22,decay23,decay24, decay25,decay26,decay27,decay28,
decay29,decay30,decay31,decay32,decay33,decay34}.
The scale of the mass
then provides a natural cut-off scale $m_D/2$ for $e^\pm$ energy spectrum. To explain FERMI-LAT and HESS data, the DM mass is then required to be around 3 TeV. The typical
life-time required to fit data is a few times of $10^{26}$s. This time is much longer than the life-time of the universe and will not cause other cosmological problems.

One, of course, should not exclude the possibility that there may be
other explanations~\cite{lorentz,other-exp,
other-exp1,other-exp2,other-exp3,other-exp-new,other-exp4,other-exp5, other-exp6,
other-exp7,other-exp8,other-exp9,other-exp10,other-exp11}
One of such possibilities is that the $e^\pm$ excess is produced by near by pulsars.
Electrons in the intense rotating magnetic field that surrounds the neutron star can emit synchrotron radiation that is energetic
enough to produce electron and positron pairs, but much harder to produce proton and anti-proton pairs. This provides a nice answer
to why PAMELA only observed positron but not anti-proton excesses. The resulting positron spectrum can be modeled as a product of a power
law and a decaying exponential with a cut-off in energy. This explains why the spectrum falls off at higher energies.

In this brief review, we will concentrate on discussions of how DM annihilation can explain $e^\pm$ excess in cosmic ray energy spectrum.
The review is arrange as the following. In Sec. II, we review propagation mechanism relating the sources of $e^\pm$ and the detected spectrum.  In Sec. III. we discuss
$e^\pm$ excesses from DM annihilation. Both Sommerfeld and Briet-Wigner mechanisms for a large boost factor will be explained in some details.
In Sec. IV some model building aspects of DM will be discussed.  Finally in Sec. V we give our conclusions.

\section{The cosmic $e^\pm$ spectrum and the boost factor}

The detected spectrum of $e^\pm$ on the earth are different than that of the spectrum produced at the sources. The propagation from the sources to the point of detection will distort the shape because charged particles propagate diffusively in the galaxy. Some of the main effects affecting the spectrum are the interactions with interstellar media when going through galactic turbulent magnetic field and radiation which lead to energy losses of the propagating particles, and the overall convection driven by the galactic wind and re-acceleration due to the interstellar shock waves. It is a non-trivial matter to get reliable estimate for $e^\pm$ energy spectrum taking all effects into account. Nevertheless theoretical efforts have been made to estimate the background and DM signal. A commonly used numerical package is the GALPROP which takes into account many known astrophysics effects of our galaxy~\cite{calprop}. This package can provide many details for energy spectra of $e^\pm$, anti-proton and other particles. To have some understanding of the physics involved and also to have some simple estimate of the $e^\pm$ spectrum, we will describe a simplified method to evaluate the $e^\pm$ spectrum from DM annihilation in the following.

Neglecting convection and re-acceleration effects, the flux per unit energy of ultra-relativistic positron or electron is given by $\Phi_{e}(t,r,E) = f(t,r,E)/4\pi$ with $f$ obeys the diffusion equation~\cite{strumia}
\begin{eqnarray}
{\partial f\over \partial t} = K(E) \bigtriangledown^2 f + {\partial (b(E) f)\over \partial E} + Q_e\;,
\label{df}
\end{eqnarray}
where $K(E)$ is the diffusion coefficient which are usually parameterized as $K_0(E/\mbox{GeV})^\delta$, and $b(E) = E^2/(\mbox{GeV}\tau_E)$ is the energy loss coefficient with $\tau_E = 10^{16}$s. These terms describe transport through the turbulent magnetic fields and energy loss due to synchrotron radiation and inverse Compton scattering on galactic photons. $Q_e$ is the source term given by
\begin{eqnarray}
&& Q_e(\vec r, E) = {1\over 2} \left ( { \rho(\vec r)\over m_D}\right )^2 <\sigma v> {dN_e\over d E}\;,
\end{eqnarray}
where $dN_e/dE$ is the spectrum of the electron or positron produced by DM annihilation, and $\rho(\vec r)$ is DM
density at $\vec r$.

The above equation is then solved in a diffusive region with the shape of a solid
cylinder that sandwiches the galactic plane, with height 2L in the z direction and radius R = 20 kpc in the
r direction. The location of the solar system corresponds to $\vec r  = (r_{sun}; z_{sun}) = ((8.5\pm 0.5) kpc; 0)$.
The boundary conditions are usually set to be that the $e^+$ or $e^-$ density vanishes on the surface of the
diffusive cylinder, outside of which turbulent magnetic fields can be neglected so that positrons
freely propagate and escape.

The values of the propagation parameters $\delta$, $K_0$ and L are deduced
from a variety of cosmic ray data and modelizations. The following three sets of parameters have been used frequently~\cite{parameters-astro}:
i) $Min:\;\;\delta = 0.55$, $K_0 = 0.000595 \mbox{kpc}^2/Myr$, $L = 1 \mbox{kpc}\;$;
ii) $Med:\;\;\delta = 0.70$, $K_0 = 0.0112 \mbox{kpc}^2/Myr$, $L = 4 \mbox{kpc}\;$;
and iii) $Max:\;\;\delta = 0.46$, $K_0 = 0.0765 \mbox{kpc}^2/Myr$, $L = 15 \mbox{kpc}\;$.

To finally obtain the solution for eq.(\ref{df}) one has also to know the details of the DM halo profile.  There are different models. Some of the popular ones
can be casted into the following form:
\begin{eqnarray}
\rho(\vec r) = \rho_{sun} \left ({r_{sun}\over r}\right )^\gamma \left ( {1+(r_{sun}/r_s)^\alpha\over 1+(r/r_s)^\alpha}\right )^{(\beta-\gamma)/\alpha}\;,
\end{eqnarray}
where $\rho_{sun}$ is the DM density at the earth position which is believed to be in the range $0.2 \sim 0.7$ and most of the studies use 0.3. For the
other parameters $\alpha$, $\beta$, $\gamma$ and $r_s$ in the profile, different values have been used. For example: a) the Core Isothermal (CI) model~\cite{ci} has
$\alpha = 2$, $\beta = 2$, $\gamma = 0$ and $r_s = 5 kpc$; b) the Navarro, Frenk and White (NFW) model~\cite{nfw} has $\alpha = 1$, $\beta = 3$, $\gamma = 1$ and $r_s = 20 kpc$; and c) the Moore model~\cite{more} has $\alpha = 1$, $\beta = 3$, $\gamma = 1.16$ and $r_s = 30 kpc$.

For practical uses, it is convenient to factor out the galactic astrophysics modeling for the propagation and particle physics modeling of the inject spectrum $dN_e/dE$, and write the final flux of stationary solution of eq.(\ref{df}) ($\partial f/\partial E = 0$) at the detection point in the following form
\begin{eqnarray}
&&\Phi_e(E, \vec r_{sun}) = B{1\over 4 \pi b(E)}{1\over 2} \left ({\rho_{sun}\over m_D}\right )^2 \int^{m_D}_E dE' <\sigma v> {dN_e\over d E'} I_a(\lambda_D(E,E'))\;,
\label{phi-form}
\end{eqnarray}
where $\lambda_D^2 = 4 K_0 \tau_E ((E'/\mbox{GeV})^{\delta -1} - (E/\mbox{GeV})^{\delta -1})/(\delta -1)$. The function $I_{a}$ encodes the galactic astrophysics. The particle physics producing the inject positron is contained in $dN_e/dE$. Numerical solutions for $I_{a}$ have been obtained in Ref. \cite{strumia} using the CI, NFW and More models for the Min, Med and Max parameter sets. The factor $B$ is the so called boost factor. If the model of propagation is correct, the factor $B$ should be equal to 1. Approximate analytic forms for $\Phi_e$ have also been obtained in the literature. For example, for the NFW profile,
the annihilation flux, to a good approximation one can write $I_a$ in the following form~\cite{strumia}
\begin{eqnarray}
I(\lambda_D) = a_0 + a_1 tanh\left ({b_1-l\over c_1}\right )\left (a_2 exp\left [-{(l-b_2)^2\over c_2}\right ] + a_3\right )\;,
\end{eqnarray}
where $l = log_{10} (\lambda_D/kpc)$.

Fitting numerical results, the following are obtained in Ref. \cite{strumia}
\begin{eqnarray}
Min\;:&&a = -0.9716\;,\;b =-10.012 \;;\nonumber\\
&&a_0=0.500\;,\;a_1 = 0.774\;,\;a_2 = -0.448\;,\;a_3 = 0.649\;,\nonumber\\
&&b_1 = 0.096\;,b_2 = 192.8\;,c_1 = 0.211\;,c_2 = 33.88\;.
\end{eqnarray}
\begin{eqnarray}
Med\;:&&a= -1.0203\;,\;b=-1.4493\;;\nonumber\\
&&a_0=0.502\;,\;a_1 = 0.621\;,\;a_2 = 0.688\;,\;a_3 =0.806\;,\nonumber\\
&&b_1 = 0.891\;,b_2 = 0.721\;,c_1 = 0.143\;,c_2 = 0.071\;.
\end{eqnarray}
\begin{eqnarray}
Max\;:&&a = -0.9809\;,\;b = -1.1456\;;\nonumber\\
&&a_0=0.502\;,\;a_1 = 0.756\;,\;a_2 = 1.533\;,\;a_3=0.672\;,\nonumber\\
&&b_1 = 1.205\;,b_2 = 0.799\;,c_1 = 0.155\;,c_2 = 0.067\;.
\end{eqnarray}

To compare with data, one has also to have knowledge about the background. The background $e^\pm$ fluxes from astrophysical sources are believed to be
mainly due to supernova explosions for the primary electrons and from the interactions
between the cosmic ray nuclei, such as  proton and light atoms, such as hydrogen
and helium, in the interstellar medium for the secondary electrons and positrons. They are
commonly parameterized in the following form~\cite{background},
\begin{eqnarray}
&&\Phi_{e^-}^{bkgd, prim} = {0.16 E^{-1.1}\over 1+11 E^{0.9} + 3.2 E^{2.15}}\;,\;\;\;\; \Phi_{e^-}^{bkgd,sec} = {0.7 E^{0.7}\over 1 + 110 E^{1.5} + 580 E^{4.2}}\;,\nonumber\\
&&\Phi^{bkgd,sec}_e = {4.5 E^{0.7}\over 1 + 650 E^{2.3} + 15000E^{4.2}}\;.
\end{eqnarray}
In the above the energy $E$ is in unit GeV.

With the background $e^\pm$ spectra known, one can say more about the role of DM in explaining the observed $e^\pm$ excesses.
%

The above set of background spectra agree well with the more sophisticated numerical simulation results shown in Figs. \ref{data-pamela} and \ref{data-fermi} for energy larger than 10  GeV. The $e^\pm$ excesses observed by
PAMELA, ATIC, FERMI-LAT and HESS have to be due to other contributions. If DM annihilation can explain the excesses,
the ratio $\Phi_{e^+}/(\Phi_{e^+} + \Phi_{e^-})_{data}$ observed by PAMELA, and the normalized flux $E^3_e(\Phi_{e^+} + \Phi_{e^-})_{data}$ observed by
ATIC, FERMI-LAT and HESS must be equal to the background plus the DM generated fluxes,
\begin{eqnarray}
&&\left ({\Phi_{e^+}\over \Phi_{e^+} + \Phi_{e^-}}\right )_{data} = {\Phi^D_{e^+} + \Phi^{bkgd,sec}_{e^+}\over \Phi^D_{e^+} + \Phi_{e^+}^{bkgd,sec} + \Phi^D_{e^-} + \Phi_{e^-}^{bkgd,sec} + \kappa \Phi_{e^-}^{bkgd,prim}}\;,\\
&&\nonumber\\
&&E^3_e(\Phi_{e^+} + \Phi_{e^-})_{data} = E^3_e (\Phi^D_{e^+} + \Phi_{e^+}^{bkgd,sec} + \Phi^D_{e^-} + \Phi_{e^-}^{bkgd,sec} + \kappa \Phi_{e^-}^{bkgd,prim})\;.\nonumber
\end{eqnarray}
Note that in the above a parameter $\kappa$ has been introduced in
the equations to take care uncertainties of primary background $e^-$
production. By adjusting $\kappa$ one expects to make a better fit
to simulation data and is usually taken to be~\cite{ kappa,kappa1}
0.8. If there is no DM contribution, the data show a large $e^\pm$ excesses. DM annihilation is one of the most interesting possibilities. If true the $e^\pm$ excess data then determine how DM contribute to the cosmic spectrum through $dN_e/dE$.
%

For DM annihilation, the parameters involved regarding DM properties
are the DM mass $m_D$, the annihilation rate $<\sigma v>$, and the
spectrum $dN_{e}/dE$ from annihilation. Since the excesses go up to
the TeV region, $m_D$ must also be in the TeV region to cover the
full range of excesses. With DM mass fixed, one needs to worry about
what type of final states in the DM annihilation can produce the
observed $e^\pm$ spectrum shape. This depends on the properties of
DM that is responsible for $dN_{e}/dE$. Several model independent
studies have been carried
out~\cite{strumia1,model-independent,m-ind1,m-ind2,m-ind3,m-ind4,m-ind5,
m-ind6,m-ind7, m-ind8}
and find that the shape of the spectrum can be easily obtained by several types of
annihilation final states which we will comment on later.

If one uses the annihilation rate determined from relic DM calculation to calculate
the $e^\pm$ excesses, one finds that the resulting excesses from the about propagation model calculations
would be smaller by a factor of 100 to 1000.
A large boost factor $B$ is needed to explain the data. This is an important issue
needs to be addressed before a consistent model can be constructed and tested further.
%

This problem may be due our lacking of knowledge of DM substructures
in our galaxy. Detailed calculations based on the N-body simulation
shows that the boost factor from DM substructures is generally less
than~\cite{xiao-jun2,xiao-jun21,xiao-jun22,xiao-jun23}
$\sim 10$.  Several other mechanisms have been proposed, including
DM non-thermal production
mechanism~\cite{nonthermal,nonthermal1,nonthermal2}
the Sommerfeld
effect~\cite{strumia1}\cite{arkani}\cite{nojiri,nojiri1,nojiri2,s-new,sommerfeld-other,sommerfeld-other1}
and the Breit-Wigner resonance enhancement
effect~\cite{bi-he}\cite{breit,breit1}.
The non-thermal production mechanism is to detach the relic DM density from the annihilation rate
producing the $e^\pm$ excesses. The annihilation rate is taken to be a parameter to be determined
by the $e^\pm$ excess data. $B = 1$ can produce large enough excess. On the other hand,
the Sommerfeld and Breit-Wigner mechanisms start with the annihilation rate determined by
relic DM density and dynamically determine the boost factor. We consider these latter two
effects to be more natural. In the follow section we discuss these two mechanisms.

\section{The Sommerfeld and Breit-Wigner enhancement factors}

\subsection{The Sommerfeld enhancement factor}

The Sommerfeld enhancement is a non-relativistic quantum mechanical effect~\cite{sommerfeld}. Since at the epoch
of relic DM got out of thermal equilibrium, the DM are non-relativistic, one can teat the problem with
non-relativistic Schrodinger quantum mechanics. A large boost factor can be produced if DM interacts with a
light particle.

The annihilation of DM is usually a short distance effect.  Assuming that it happens at the origin of a coordinate. The annihilate cross section can be approximated by a potential of the form $V_a\delta(\vec r)$. Imagining  now that the DM is moving in the z-direction, the wave-function can be written up to some overall normalization factor as, $\psi^0_k (\vec r) = e^{ikz}$. It is obvious that the annihilation cross section is proportional to $|\psi^0_k(0)|^2$ since this factor represents the DM density at the origin where the annihilation occurs. The cross section $\sigma_0$ due to the short distance interaction can be easily obtained by the usual scattering theoretical calculations.

If in addition to the short distance interaction, there is an exchange of a massless particle between the two annihilation DMs, a long range interaction potential of the form $V_C = - \alpha/r$ between the DM will be generated. Before the annihilation happens, the
$1/r$ long range force, as is well known, will distort the DM wave-function $\psi(\vec r)$ and therefore modify the DM density at the origin. The cross section is then proportional to
$|\psi_k(0)|^2$ resulting in a modification factor, $S = |\psi_k(0)|^2/|\psi^0_k(0)|^2$ and the cross section  is given by
$\sigma = \sigma^0 S$. This factor is the Sommerfeld factor.

The S factor due to a long range Coulomb like potential $V_C(r) = - \alpha/r$ can be obtained by solving the Schrodinger equation
\begin{eqnarray}
E\psi_k(\vec r) = (-{1\over 2 \mu}\bigtriangledown^2 + V_C(r))\psi_k(\vec r )\;,
\end{eqnarray}
where $E = k^2/2\mu > 0$. $\mu = m_D m_D/(m_D + m_D)= m_D/2$ is the reduced mass in the center of mass frame of the two annihilating DM.

This equation is a standard central force problem and can be solved by separation of variable in the form $\psi_k(\vec r) = \sum_{l,m} A_{lm}R_l(r) Y^m_l(\theta, \phi)$ with $Y^m_l(\theta, \phi)$ being the spherical harmonic function. $R_l(r)$ satisfies
\begin{eqnarray}
\left ({1\over r^2} {d\over dr} r^2 {d\over dr} + k^2 - 2\mu  (V_C(r) + {l(l+1)\over r^2})\right ) R_l(r) = 0\;.\label{shrodinger}
\end{eqnarray}

To obtain an analytic solution suitable for scattering problem, let us write $R_l(r) = r^l e^{ikr} f_l(r)$. The function $f_l(r)$ satisfies the confluent hypergeometric equation~\cite{landau}
\begin{eqnarray}
z {d^2\over dz^2} f_l(z) + (2(l+1) - z) {d\over dz } f_l(z) - (l+1 + in) f_l(z) = 0\;,
\end{eqnarray}
where $z = - i k r$ and $n = - \mu \alpha/k$. The solution is the regular confluent function $f_l(z) = {}_1F_1(l+1+in, 2(l+1), z)$ since at $r=0$, the wave function must be finite. Then
\begin{eqnarray}
R_l(r) = C_l r^l e^{ikr}  {}_1F_1(l+1+in, 2(l+1), -2ikr)\;,
\end{eqnarray}

To determine the constant $C_l$, one matches the asymptotic behavior at $r\to \infty$ of
the partial wave decomposition of a plane wave scattered by a potential
\begin{eqnarray}
\psi_k(r,\theta) &\to& e^{ikz} + {e^{ikr}\over r} \sum_0^\infty (2l+1) {e^{2i\delta_l} - 1\over 2 i k r} P_l(\cos\theta)\nonumber\\
 &=& {1\over 2ik r }\sum_0^\infty (2l+1) \left ( e^{ir + 2i\delta_l} - e^{-i(kr - l\pi}\right )P_l(\cos\theta)\;,\label{ppp}
\end{eqnarray}
with the behavior of $\psi_k(r,\theta) = \sum_l C_l R_l(r)P_l(\cos\theta)$ at $r\to \infty$
\begin{eqnarray}
&&\sum_l C_l {e^{n\pi/2} \Gamma(2l+1)\over (2 k)^l \Gamma(l+1+in)} {1\over 2ikr} \left ( e^{i(kr - l\pi/2 - n\ln(2kr) + \eta_l} - e^{- i(kr - l\pi/2 - n\ln(2kr) + \eta_l} \right )\nonumber\\
&&\;\;\;=  C_l {e^{n\pi/2} \Gamma(2l+1)\over (2 k)^l \Gamma(l+1+in)}e^{-i( l\pi/2 + n\ln 2 -\eta_l})\\\label{pp}
&&\;\;\;\times  {1\over 2ikr}  \left ( e^{i(kr  - n\ln(kr) -2n\ln 2 +  2 \eta_l)} - e^{- i(kr - n\ln(kr)-l\pi)} \right )P_l(\cos\theta)\;; \nonumber
\end{eqnarray}
where $\eta_l = arg\Gamma(l+1 + i n)$.

Naively it seems that, keeping $C_l$ independent of $r$, it is not possible to written the asymptotic form of eq.(16) into the corresponding coefficient (the same $l$) in eq. (\ref{ppp}) because the $\ln(k r)$ factor in the exponential. This is a well known fact which happens to the Coulomb potential scattering. The $\ln (kr)$ needs to be kept in the exponential due to the long range interaction nature of the potential. In this case the scattering phase $\delta_l$ is equal to $\eta_l - n\ln2$, and $C_l$ is then given by
\begin{eqnarray}
C_l = {1\over (2l)!} (2i k)^l \Gamma(l+1+in) e^{-n\pi/2}e^{i(n\ln 2 - \eta_l)}\;.
\end{eqnarray}

Note that whatever modification to the wave-function at $r=0$, it only happens to the S-wave, $l=0$ case, since for $l\neq 0$, $R_l(r) \sim r^l e^{ikr} {}_1F_1(l+1+in, 2(l+1),-2ikr)$ is zero.  One immediately finds that
\begin{eqnarray}
\psi_k(0) =\Gamma(1+in) e^{-n\pi/2} e^{i(n\ln2 - \eta_0)}\;.
\end{eqnarray}

Using the identity $\Gamma(1+x) \Gamma(1-x) = x\pi/\sin(x\pi)$, one finally obtains
\begin{eqnarray}
S = {|\psi_k(0)|^2 \over |\psi^0_k(0)|^2 }=  {2n\pi\over e^{2n\pi} - 1}   ={ -\alpha\pi/v\over e^{-\alpha \pi/v} -1}\;.
\end{eqnarray}

With the Sommerfeld factor included, the cross section is given by
\begin{eqnarray}
\sigma(v) v = \sigma^0(v)v {- \alpha\pi/v\over e^{- \alpha \pi/v} -1}\;.
\end{eqnarray}
If $\alpha$ is positive, corresponding to an attractive potential, $S$ is an enhancement factor. While for a negative $\alpha$,
corresponding to a repulsive potential, $S$  is a suppression factor. Also it is velocity dependent. $S$ plays the role of the boost factor.

For S-wave annihilation, if the DM is not close to a resonant region, the cross section is almost a constant in DM velocity. Therefore boost factor
for DM annihilation resulting from the relic DM is approximately the ratio of the DM velocity $v_r$ at the relic DM decoupling time in the early universe and the velocity $v_p$ of DM in our galaxy halo at the present, that is $B\sim v_r/v_p$.
The average $v_r$ at the decoupling temperature  $T_d$ is about $\sqrt{2 T_d /m_D}$ with  $T_d/m_D \sim 1/20$ leading to  $v_r \approx 0.3$. Model estimates show that $v_p$ is about a few times $10^{-4}$. With these numbers, one can see that the Sommerfeld enhancement can easily produce the needed large boost factor.

\begin{figure}[t]
\begin{center}\includegraphics[width=0.5\columnwidth
]{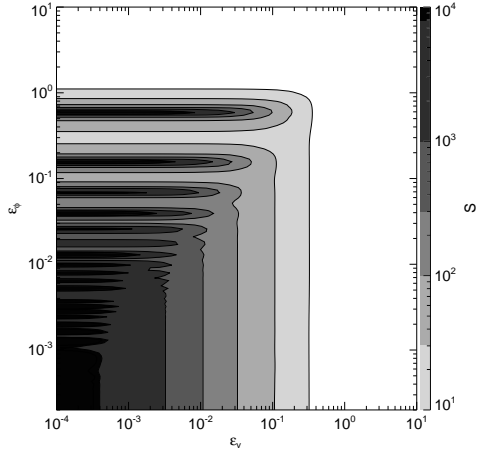}
\end{center}
\caption{The Sommerfeld factor contour as a function of $\epsilon_v = v/\alpha$ and $\epsilon_\phi = m_\phi/\alpha m_D$.}\label{sommerfeldfig}
\end{figure}

In many practical situations, the interaction between DM is not mediated by a massless particle but a massive one, and
the potential produced is a Yukawa potential $V_Y(r) = -(\alpha/r)e^{-m_\phi r}$. Here $m_\phi$ is the mass of the mediating particle. One needs to solve the differential equation in eq.(\ref{ppp}) with $V_C(r)$ replaced by $V_Y(r)$. Unfortunately with $V_Y(r)$, it is not possible to obtain a simple analytic solution. To obtain the corresponding Sommerfeld enhancement factor, numerical calculation is needed. Fig. \ref{sommerfeldfig} shows the enhancement factor as a function of model parameters obtained in Ref. \cite{arkani}. There are regions of parameters, resonant regions, where the enhancement factor can easily be as large as 1000.

One can understand some qualitative features without detailed numerical calculations. The Yukawa potential is not a long range interaction but one with a force range cut-off at $r \sim 1/m_\phi $ and therefore the enhancement factor will be smaller than that due to a Coulomb potential. For $r <  1/m_\phi$ one can expand the exponential part of the potential in power of $r$. Keeping the leading correction,
the potential is: $V_Y(r) \sim -(\alpha/r - \alpha m_\phi)$. Comparing with eq.(\ref{shrodinger}), one notices that if $v^2 >> v^2_s = 4 |\alpha| m_\phi/m_D$, the correction term can be neglected and get back to the Coulomb potential case. At the velocity $v_s$ the Sommerfeld enhancement starts to damp out. When the velocity slowed further down to where the corresponding deBrolie wave length $1/m_D v_d$ of the DM  to be comparable with the force range $1/m_\phi$, the Sommerfeld enhancement saturates itself. This sets a natural limit of the enhancement~\cite{arkani}, $\pi\alpha m_D/m_\phi$.

For a detailed discussion, an more accurate form for
the averaged annihilation rate should be used which can be
written as~\cite{gemini}
\begin{equation}
\langle \sigma v\rangle=\frac{1}{n_{EQ}^2}\frac{m_{D}}{64\pi^4 x}
\int_{4m_{D}^2}^{\infty} \hat{\sigma}(s)\sqrt{s}K_1\left(
\frac{x\sqrt{s}}{m_{D}}\right){\rm d}s,\label{avr}
\end{equation}
with
\begin{eqnarray}
n_{EQ} &=& \frac{g_i}{2\pi^2}\frac{m_{\psi}^3}{x}K_2(x)\;,\nonumber\\
\hat{\sigma}(s) &=& 2g_i^2m_{\psi}\sqrt{s-4m_{\psi}^2}\cdot \sigma v,
\end{eqnarray}
where $x = m_D/T$ and $g_i$ is the internal degrees of freedom of DM particle
which is equal to 4 for a Dirac fermion. $K_1(y)$ and $K_2(y)$
are the modified Bessel functions of the second type.

In the non-relativistic case, the above reduces to the Maxwell-Boltzmann average,
\begin{eqnarray}
\langle \sigma v \rangle  = \int \sigma(v) v \sqrt{{2\over \pi}}{1\over \delta^3} v^2 e^{-v^2/2\delta^2}\;,\label{avn}
\end{eqnarray}
where $\delta$ is the average DM velocity. At the relic DM decoupling, it is $\sqrt{2 T_d/m_D}$, and at present in our galaxy halo, it is about $10^{-4}$.

Finally to  determine the parameters one should solve the standard
Boltzmann equation governing the DM abundance~\cite{gemini}
\begin{eqnarray}
{dY\over dx} = - {xs(x)\over H} <\sigma v> (Y^2- Y^2_{EQ})\;,
\end{eqnarray}
where $Y = n/s(x)$ with $n$ the DM number density and $s(x) = 2\pi^2 g_* m^3_D/45 x^3$  the entropy density.
$H = \sqrt{4\pi^3/45}m^2_D/M_{PL}$ is the Hubble constant evaluated at $x= 1$. Here $M_{PL}\approx 1.22 \times 10^{19}$ GeV
is the Planck mass. $g_*$ is the total relativistic degrees of freedom. In the SM $g_* = 106.75$. The $Y$ value at thermal equilibrium $Y_{EQ}$ is given by: $(45/4\sqrt{2}\pi^{7/2})(g_i/g_*)x^{3/2}e^{-x}$.

The rest challenge is more a particle physics one, finding a particle physics model which
can have a particle $\phi$ with small mass and interact with DM to produce the required boost factor.

\subsection{The Briet-Wigner resonant enhancement factor}

The Breit-Wigner enhancement mechanism can produce a large boost factor~$^{15,143,144}$
if the annihilation of DM is through exchange of
a particle in the S-channel with mass close to $2m_D$. Let us consider an example that the DM annihilates into lepton pairs by exchange of a S-channel $Z^\prime$,
$\psi \bar \psi \to Z^\prime \to l \bar l$ through the following Lagrangian
\begin{eqnarray}
L = (a g^\prime \bar \psi \gamma^\mu \psi + g^\prime \bar l \gamma^\mu l) Z^\prime_\mu\;.
\end{eqnarray}

The interaction rate, to the leading order in velocity $v$, is given by~\cite{bi-he}
\begin{eqnarray}
\sigma(v) v ={1\over \pi} {a^2g^{\prime 4}m^2_D\over (s-m^2_{Z^\prime})^2 + \Gamma^2_{Z^\prime} m^2_{Z^\prime}}\;,
\end{eqnarray}
where $\Gamma_{Z^\prime}$ is the total $Z^\prime$ decay width and $s$ is the center of mass frame energy squared.

If $2m_D$ is far away from $m_{Z^\prime}$ the interaction rate is almost a constant. After the parameters fixed by relic DM density requirement,
the interaction rate is constrained to be too small to explain the data. A large boost factor is needed.
The boost factor
can arise from the fact that when the $Z'$ mass is close to $2 m_D$, the
annihilation rate is close to the resonant point. In this case the interaction rate is very sensitive to velocity of the DM.
To see this let us rewrite the above annihilation rate as
\begin{eqnarray}
\sigma(v) v  = {a^2 g^{\prime 4}\over 16 \pi m^2_D} {1\over (\delta + v^2/4)^2 + \gamma^2}\;,
\end{eqnarray}
where we have used the non-relativistic limit of
$s = 4 m^2_D + m^2_D v^2$, with $\delta$ and $\gamma$
defined as $m^2_{Z^\prime} = 4 m^2_D (1-\delta)$, and $\gamma^2
= \Gamma^2_{Z^\prime}(1-\delta)/4 m^2_D$.

It is clear that for small enough
$\delta$ and $\gamma$, the annihilation rate is very sensitive to
the velocity $v$. At
lower velocity, the annihilation rate is
enhanced. This results in
a very different picture of DM annihilation than the case
for the usual non-resonant annihilation where the annihilation
rate is not sensitive to DM velocity. The
annihilation process does not freeze out even after the usual
``freeze out'' time in the non-resonant annihilation case due to the
enhanced annihilation rate at lower energies in the early universe. To produce the
observed DM relic density, the annihilation rate at zero
temperature is required to be larger than the usual one, and
therefor a boost factor. With appropriate $\delta$ and $\gamma$, a
large enough boost factor $B$ can be produced.

\begin{figure}[t]
\begin{center} \includegraphics[width=0.5\columnwidth
]{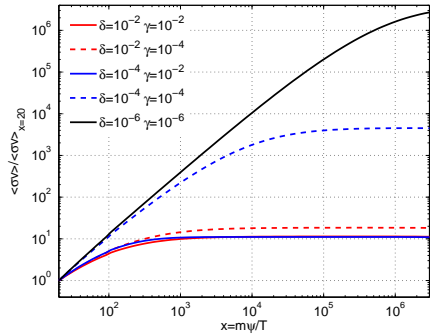}
\end{center}
\caption{The Breit-Wigner enhanced relative interaction rate
$\langle\sigma v\rangle/\langle\sigma v\rangle_{x=20}$ as a
function of time $x$. $m_\psi$ here is $m_D$.\label{boost}}
\end{figure}

Once the interaction rate is obtained, one can use the formulae in eqs.(\ref{avr}) and (\ref{avn}) to obtain the boost factor
$B$. A numerical evaluation obtained in Ref.\cite{bi-he} for the Breit-Wigner enhancement is shown in Fig. \ref{boost}.

\section{Theoretical models for dark matter }

\subsection{Dark matter models}

Many candidate WIMP DM models have been studied in the context of
the recent $e^\pm$
data~\cite{arkani}\cite{leptophilic,lep1,lep2,lep3,bi-he}\cite{susy,susy1,susy2,susy3,susy4,susy5,susy6,susy7,
susy8,susy9,susy10,susy11,susy12,susy13,susy14,
susy15,susy16,susy17}\cite{nmssmhooper,nmssmhooper1,nmssmlykken}\cite{
kk,kk1,kk2}\cite{light, light1,light2, light3,light4,light5,
light6,light7,light8, light9,light10,light11,
light12,light13,light14, light15,light16,light17,
light18,light19,light20, light21,light22,light23}\cite{annihilation,
annihilation1, annihilation2, annihilation3, annihilation4,
annihilation5, annihilation6, annihilation7, annihilation8,
annihilation9, annihilation10, annihilation11, annihilation12,
annihilation13, annihilation14, annihilation15,annihilation16}.
But not all of them can explain all the features of the $e^\pm$
excess in cosmic ray. It turns out that the spectrum shape of the observed $e^\pm$ excesses
is easy to reproduce. For example, two boy annihilation directly into $e^\pm$ pair produces a
hard spectrum which can be made consistent with ATIC data, but disfavored by FERMI-LAT and HESS
data. If the two body final states are $\mu^\pm$ or $\tau^\pm$ pair, their subsequent decays
into $e^\pm$ can produce a soft enough spectrum consistent with FERMI-LAT and HESS data.
The large boost factor and $e^\pm$ excesses in energy spectrum but not in anti-proton spectrum
required from the data eliminate many candidate DM annihilation models.
There are some doubts on the reliability of the null result for the anti-proton. But before
further experimental results disprove this result, one cannot simple ignore it. In the following
discussions, we will take these two as requirements for a successful model.

The simplest DM model is the Darkon model~\cite{darkon}. This model
is the simplest in the sense that it needs the least extension from
the SM. It contains in addition to the SM particles,  just one real
SM singlet scalar. This singlet scalar plays the role of DM. The
annihilation is through exchange of the SM Higgs boson. The mass can
be as large as  a TeV, but should not be too small (less than a GeV)
in order not to produce too large detection cross section to be
ruled out by data~\cite{he-darkon,he-darkon1}
Although large boost factor can
be induced if the Higgs mass is close to 2 times of the darkon mass
through the Breit-Winger enhancement mechanism, it produces too many
anti-protons in the cosmic ray after fitting the $e^\pm$ excess
data. This is  because that in this model the couplings of Higgs to
SM fermions is proportional to fermion masses, the annihilation
tends to favor heavy final states. Further extension of the model is
needed to explain data.

The most popular DM candidate is the lightest supersymmetric
particle (LSP) in the minimal supersymmetric standard model (MSSM).
The MSSM LSP has been proposed to explain the $e^\pm$
excesses~\cite{susy,susy1,susy2,susy3,susy4,susy5,susy6,susy7,
susy8,susy9,susy10,susy11,susy12,susy13,susy14,
susy15,susy16,susy17}
However the MSSM also has problem. The LSP is a linear
combination of photino, zino and higgsino. It usually has a large hadronic annihilation
fraction in conflict with the PAMELA data on anti-proton cosmic ray. Also in this model
there is the problem to realize a large boost factor. Gravitino DM candidate also has
similar problems. Extensions are needed. There are a few papers on this subject.
The next MSSM (NMSSM) model has all the right ingredients to explain the data.
We will describe it more later.

In universal extra dimension (UED) models, all SM particles live in
extra dimensions. After compactification of the extra dimensions,
there are Kaluza-Klein (KK) excitations generated. If the size the
extra dimension $R$ is of order $1/TeV$. The KK mode can paly the
role of DM to explain the $e^\pm$ excesses if a discrete symmetry is
applied to make KK mode stable~\cite{ kk,kk1,kk2}.
The KK DM can annihilate into SM particles. For example, the
$U(1)_Y$ gauge boson KK mode can annihilate through t-channel
fermion KK mode into SM fermions, and produce $e^\pm$ excesses in
cosmic ray~\cite{ kk,kk1,kk2}.
However, they usually have a large hadronic fraction making the
model troublesome with PAMELA anti-proton data. This problem can be
remedied in the split-UED model where it is possible to split the
lepton and quark KK masses such that the anti-proton cosmic ray is
suppressed by larger quark KK mode masses~\cite{kk2}.
However, these models all have problem to have a large boost factor.

Many other models~\cite{arkani}\cite{
leptophilic,lep1,lep2,lep3,bi-he}\cite{light, light1,light2,
light3,light4,light5, light6,light7,light8, light9,light10,light11,
light12,light13,light14, light15,light16,light17,
light18,light19,light20, light21,light22,light23}\cite{annihilation,
annihilation1, annihilation2, annihilation3, annihilation4,
annihilation5, annihilation6, annihilation7, annihilation8,
annihilation9, annihilation10, annihilation11, annihilation12,
annihilation13, annihilation14, annihilation15,annihilation16}
have been proposed to explain the data. There are basically two
classes of models: a) kinematically limited light particle decay
models~\cite{arkani}\cite{light, light1,light2,
light3,light4,light5, light6,light7,light8, light9,light10,light11,
light12,light13,light14, light15,light16,light17,
light18,light19,light20, light21,light22,light23};
and b) leptophilic DM models~\cite{
leptophilic,lep1,lep2,lep3,bi-he}
The light particle decay model a)
requires the existence of a light particle with a mass less than the
sum of proton and anti-proton masses and thus the light particle
decays predominantly into final states containing an $e$ and/or a
$\mu$. If final states with such a particle is the dominant
annihilation channel, the $e^\pm$ produced will be able to produce
the $e^\pm$ excesses with appropriate mechanism to acquire a large
boost factor. Exchange of a light enough particle between DM can
produce a large boost factor through Sommerfeld enhancement
mechanism. The Beit-Wigner mechanism can also supply the boost
factor. The option b), leptophilic model, can be realized by
interaction of DM being leptophilic (or hadrophobic) such that
interactions of the particle mediating DM annihilate only have non-zero couplings to
leptons at tree level. In this case the mediating particle does not
have to be light. Depending on how the annihilation occurs, it is
also possible to have Sommerfeld or Brei-Wigner enhancement
mechanism. In the following we discuss two models to illustrate how
consistent models can be constructed.

\subsection{A leptophilic $Z^\prime$ model}

One of the following global symmetries in the SM can be gauged
without gauge anomalies~\cite{zp-model,
zp-model1,zp-model2,zp-model3,zp-model4}
\[
a)\;\; L_e - L_\mu , \ \ b)\;\; L_e - L_\tau , \ \ c)\;\; L_\mu - L_\tau\ .
\]
At the tree-level the $Z^\prime$ only couples to one of the pairs
$e$ and $\mu$, $e$ and $\tau$, and $\mu$ and $\tau$.  If the
$Z^\prime$ in one of these models is the mediating DM annihilation,
the resulting final states are mainly leptonic states which can lead
to excesses in $e^\pm$ observed in cosmic rays. Of course it
requires that the $Z^\prime$ to couple to
DM~\cite{strumia}\cite{lep1}\cite{bi-he}.
We assume that the DM field
is a fermionic field $\psi$ with a non-trivial $L_i - L_j$ number $a$. To have an anomaly free theory, this DM field should have vector-like coupling to
$Z^\prime$.  The
$Z^\prime$ boson can develop a finite mass $m_{Z^\prime}$ from spontaneous
$U(1)_{L_i-L_j}$ symmetry breaking by a non-zero vacuum expectation values $v_s$ of a scalar $S$ with a
non-trivial charge $L_i - L_j = b$ with $m^2_{Z^\prime} = b^2 g^{\prime 2} v_S^2$.
The $Z^\prime$ has the desired leptophilic couplings to fermions given by~\cite{bi-he}
\begin{eqnarray}
L = - g'(a \bar \psi\gamma^\mu \psi + \bar l_i \gamma^\mu l_i - \bar l_j \gamma^\mu l_j
+ \bar \nu_i \gamma^\mu L \nu_i - \bar \nu_j \gamma^\mu L \nu_j) Z^\prime_\mu\;.
\end{eqnarray}

The relic DM density is controlled
by annihilation of $\bar \psi \psi \to Z^{\prime *} \to l_i \bar l_i + \nu_i \bar \nu_i$.
The interaction rate $\sigma v$, with lepton masses neglected and summed
over the two types of charged leptons and neutrinos, is given by
\begin{eqnarray}
\sigma v = {3\over \pi} {a^2 g^{\prime 4}m^2_\psi\over (s -
m^2_{Z^\prime})^2 + \Gamma^2_{Z^\prime} m^2_{Z^\prime}}\; .\label{cs}
\end{eqnarray}
If the $Z^\prime$ mass is below the $\bar
\psi \psi$ threshold which we will assume, the dominant decay
modes of $Z^\prime$ are $Z^\prime \to \bar l_i l_i + \bar \nu_i
\nu_i$, and $\Gamma_{Z^\prime}$ is given by, neglecting lepton
masses: $\Gamma_{Z^\prime} = 3 g^{\prime 2}m_{Z^\prime}/12\pi$.

In the above expressions for $\sigma v$ and $\Gamma_{Z^\prime}$, it has been assumed that there are
only left-handed light neutrinos. If there are light right-handed
neutrinos, the factor 3 in these equations should be changed to 4.

When calculating relic DM density, one should use the above interaction rate.
However, when calculating the $e^\pm$ spectrum, the above interaction rate needs to be multiplied by a factor of 2/3
since the neutrinos in the final state do not contribute to the $e^\pm$ spectrum.

\begin{figure}[ht]
\begin{center} \includegraphics[width=0.45\columnwidth]{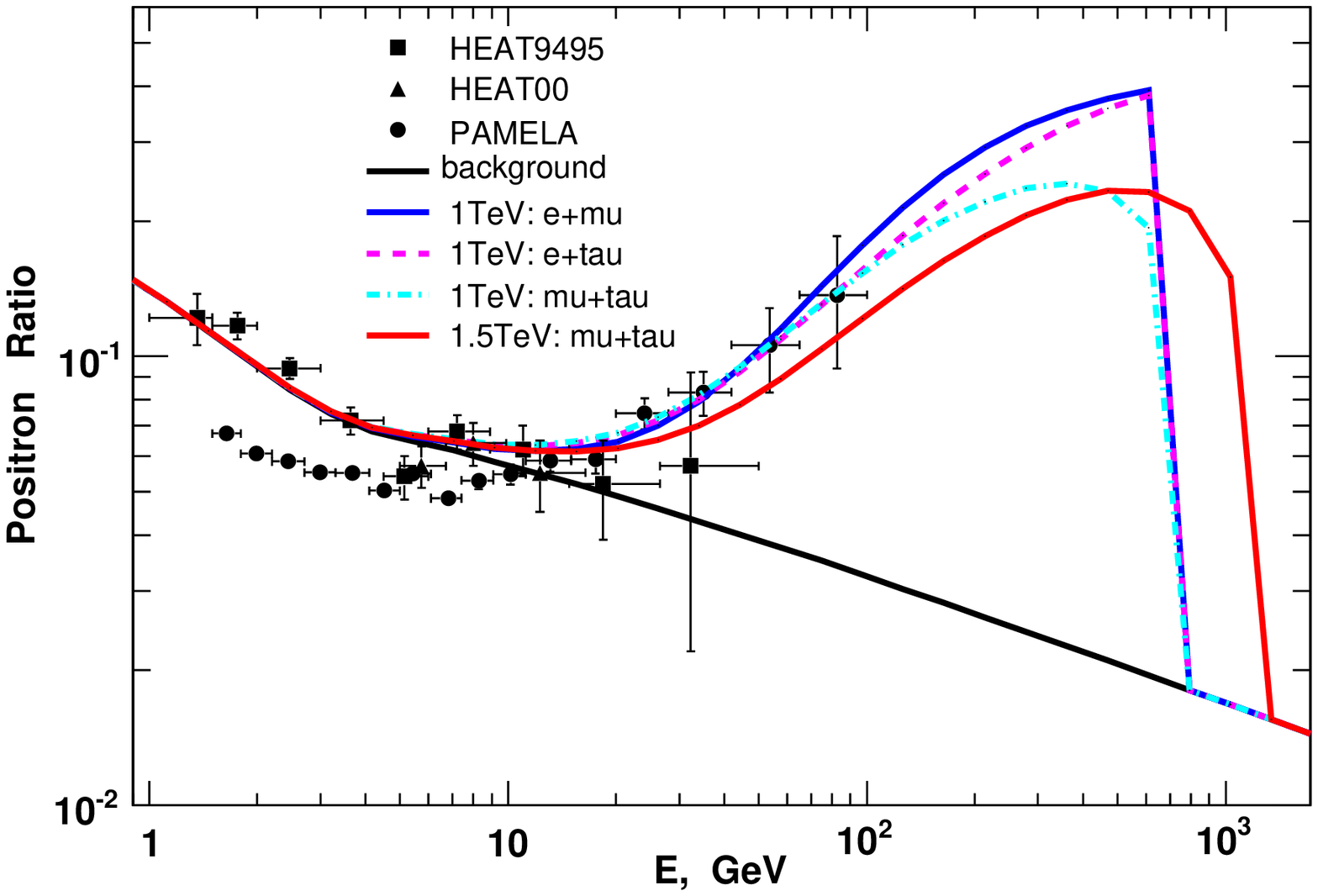}
\includegraphics[width=0.45\columnwidth]{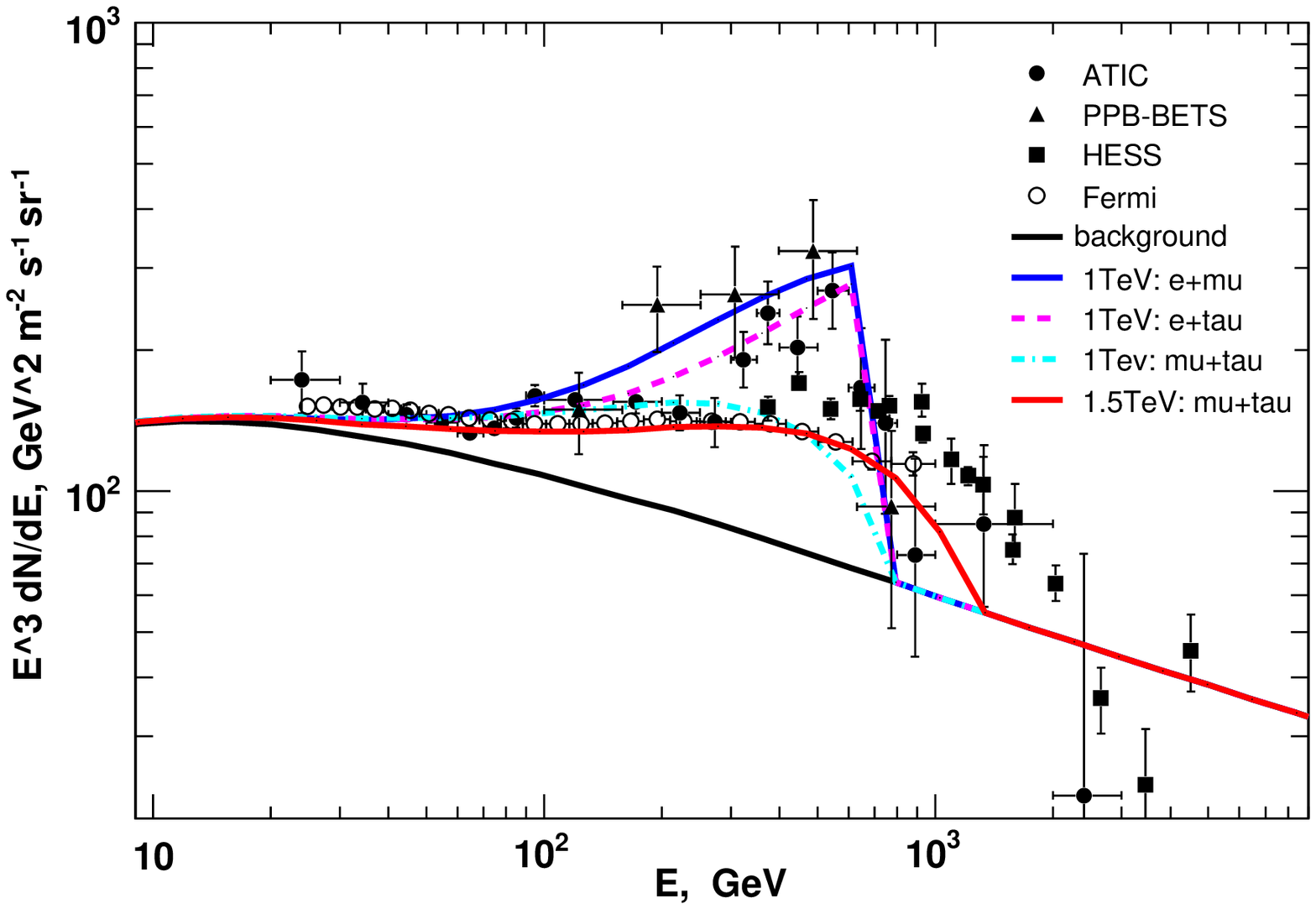}\end{center}
\caption{{\it Left:} positron fraction $e^+/(e^++e^-)$ predicted in
the $U(1)_{L_i-L_j}$ model compared with the observational data.
{\it Right:} the normalized total electron spectrum of the model,
compared with observational data. $m_D = 1.5$ TeV is used inthe
figures.}\label{elec}
\end{figure}

The Breit-Wigner enhancement factor can be used to generate the needed large boost factor by requiring $m_{Z^\prime}$ to be close to $2 m_D$.
The results obtained in Ref. \cite{bi-he} are shown in Fig. \ref{elec}. The background is calculated using
GALPROP package~\cite{calprop} with the diffusion + convection
model parameters developed in Ref. \cite{strumia}. Models a) and b), having hard $e^\pm$ in the final state, can fit ATIC and PAMELA data~\cite{bi-he} with DM mass 1 TeV.
 Model c) can fit FERMI-LAT data~\cite{bi-he}
with DM mass $1.5$  to 2 TeV. Once observational data finally settled down, some of the options can be further eliminated.

One may also construct similar model using Sommerfeld enhancement mechanism.
A possible way of achieving this is to introduce a light real scaler $\phi$ which carries no SM quantum numbers and therefore only couples to
DM in the form $\alpha \bar \psi \psi \phi$. Exchange this scalar will produce a Yukawa potential,
$V_Y(r) = - (\alpha/r) e^{-m_\phi r}$, between DM.
If the mass $m_\phi$ is small, a few GeV, there is enough parameter space where the model can fit the data. The results are similar to the case with Breit-Wigner enhancement mechanism.
The field $\phi$ can in principle mix with the SM Higgs doublet through $\phi$ and then decay to SM particles.
If the mixing is small enough it can avoid many potential phenomenological constraints.

The main difference for models with Breit-Wigner and Sommerfeld enhancement mechanisms  is that in the former model
 $m_{Z^\prime}$ is fixed to be close to $2 m_D$, while the latter model $m_{Z^\prime}$ can
have a much smaller mass leaving better chance for LHC study. Some
of the discussions on LHC physics discussed in Ref. \cite{lep1}
can
be applied. If there is kinetic mixing term~\cite{k-mixing} $F^{\mu\nu} B_{\mu\nu}$ between the field strength $B_{\mu\nu}$ of the $U(1)_Y$ and the field
strength $F^{\mu\nu}$ of the new $U(1)_{L_i-L_j}$, then there is chance to have more couplings to hadron
states making LHC study more relevant. This scenario is worth further investigation.

Another interesting possibility is that the $U(1)_{L_i-L_j}$ gauge
boson $Z^\prime$ mass is actually very small~\cite{zp-model,
zp-model1,zp-model2,zp-model3,zp-model4}
( O(GeV)). With the Sommerfeld mechanism provides the boost factor,
the $e^\pm$ is produced by t-channel $\psi \bar \psi \to Z^\prime Z^\prime$ first and then $Z^\prime$ decays into final product with $e^\pm$. In this case the light $Z^\prime$
may be produced at low energy through~\cite{zhu} $e^+e^- \to Z^\prime \gamma$.

\subsection{A light particle decay model}

As mentioned earlier that MSSM has problem accommodating $e^\pm$
excesses in cosmic ray data and there is the need to extend the
model to explain data. It has been shown that the NMSSM~\cite{nmssm,
nmssm1,nmssm2,nmssm3}
can have all the required ingredients. The NMSSM has, in addition to
the usual MSSM particle contents, a SM singlet chiral super-field
$\hat S$. The fermionic partner of $\hat S$ is the singlino, $\chi$.
The spin-0 partner $S$ contains a scalar $h = Re(S)/\sqrt{2}$ and a
pseudoscalar $a = Im(S)/\sqrt{2}$. The allowed renormalizable
super-potential $W_s$ and soft SUSY breaking potential $V_s$ are
given by~\cite{nmssmhooper,nmssmhooper1}
\begin{eqnarray}
&&W_s = v^2_0 \hat S + {1\over 2} \mu_s \hat S^2 + \lambda \hat H_u \hat H_d \hat S + {1\over 3} \kappa \hat S^3\;,\nonumber\\
&&V_s = -[{1\over 2} m_s^2 S^\dagger S + B_s S^2 + \lambda A_\lambda H_u H_d S + \kappa A_\kappa S^3] + H.C.\;,
\end{eqnarray}
where $\hat H_{u,d}$ are the MSSM SM doublet super-fields. If a $Z_3$ discrete symmetry is imposed on $W_s$ and $V_s$,
only $m_s$, $\lambda$, $\kappa$, $A_\lambda$ and $A_\kappa$ are allowed.

With suitable parameters in the model the singlino $\chi$
can have a very small mixing with other SUSY neutral fermionic fields and is the lightest super particle (LSP) playing the role of DM.
The singlino mainly annihilates into the scalar $h$ and the pseudoscalar $a$, $\chi \chi \to a h$, through $t$ and $u$ channel exchanges of $\chi$,
and s-channel exchange of $a$. With non-zero $\chi\chi h$, $\chi\chi a$ and $h aa$ couplings, $T_{h\chi\chi}$, $T_{a \chi\chi}$ and $g_{haa}$, respectively, the
interaction rate is given by~\cite{nmssmhooper}
\begin{eqnarray}
\sigma(\chi \chi \to a h) v \approx {1\over 64 \pi m^2_\chi} \left ( {1\over 16 m^2_\chi} g^2_{haa} T^2_{a\chi\chi} + T^2_{h\chi\chi} T^2_{a \chi\chi} -{1\over 2 m_\chi} g_{haa} T_{h\chi\chi} T_{a\chi\chi}^2\right ).
\end{eqnarray}
The singlino $\chi$ can also annihilate into $hh$ and $aa$ final state from $t$ and $u$ channel $\chi$ exchanges. But these contributions are
suppressed by $v^2$ and do not play a substantial role.

The $h$ and $a$ can mix with the Higgs and the pseudoscalar in the MSSM and therefore can decay into SM particles.
If
$a$ has a mass $m_a$ below the threshold of hadronization, it will not decay into proton and anti-proton pair or
even pions. If $h$ mass $m_h$ is larger than $2m_a$,
$h \to a a$ is the main decay channel for $h$. This way the final products of $\chi \chi \to a h \to a a a$ have
no proton and anti-proton complying with PAMELA data.
If $m_a$ is smaller than $2 m_\pi$ but larger than $2 m_\mu$, $a$ predominately decays into $\mu^+\mu^-$ pairs
and then $\mu \to \nu_\mu e \bar \nu_e$. In fact there is a hint from hyperCP
measurement~\cite{hyper-exp} that there is a light particle of mass 214 MeV from $\Sigma^+ \to p \mu^+\mu^-$
data and the NMSSM particle a fits that well~\cite{hypercp}.
The $e^\pm$ excess produced this way will have soft $e^\pm$ spectra. It is not possible to produce a peak
like spectrum in the ATIC data. But with $\chi$
mass $m_\chi$ of order 2 TeV, the PAMEAL and FERMI-LAT spectra can be explained.

\begin{figure}[ht]
\begin{center} \includegraphics[width=0.6\columnwidth]{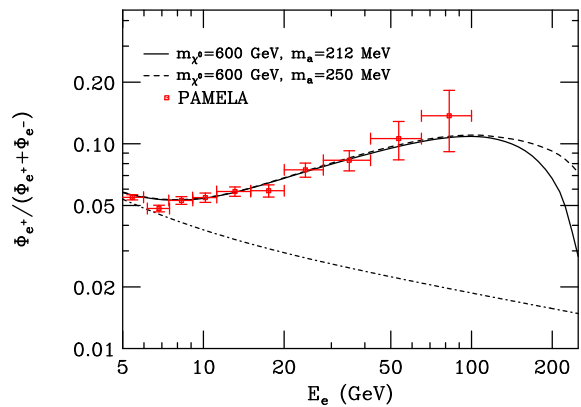}
\end{center}
\caption{The cosmic ray positron fraction resulting from singlino DM
annihilations. The $e^+$ spectrum is produced by $\chi \chi \to h a$ followed by
$h \to aa$
and $a \to \mu^+\mu^-$. $m_h$ is taken to be 10 GeV. The dot-dashed line denotes the prediction from astrophysical
secondary production alone.}\label{nmssm}
\end{figure}

Exchange of the scalar field $h$ between singlino $\chi$ produces an attractive Yukawa
potential $V_Y(r) = - (\alpha/r)e^{-m_h r}$. If $m_h$ is light enough,
less than a few GeV, a large Sommerfeld enhancement factor can be produced which then
supply the much needed boost factor. This is a quit economic
model. Fig. \ref{nmssm} shows how the PAMELA data can be fitted~\cite{nmssmhooper} with
singlino mass of 600 GeV. But with $\chi$ mass $m_\chi$ of order 2TeV, although there may be the
need of tuning the parameters, the PAMELA and FERMI-LAT spectra can be
explained.

If the mixing of $a$ with the MSSM heavy pseudoscalar $A$ is significant. The annihilation
rate $\chi \chi \to A \to a h$ can be large when the mass $m_A$ of the field $A$ is
close to $2 m_\chi$. In this case the Breit-Wigner mechanism for the large boost factor $B$
can be in operation, and also explain the data~\cite{nmssmlykken}. In this case $m_A$ is predicted to
be $2m_D$ which would in the range for 3 to 4 TeV.

In this types of model, if the light intermediate state does couple to quarks, even it is
kinematically limit to decay directly in to final states containing anti-proton or proton, there are
off-shell contributions, but the rate is suppressed.

\section{Discussions and Conclusions}

In previous discussions we have concentrated on explanation of  $e^\pm$ excesses.
Models constructed for this purpose have many other testable consequences. We list, without detailed discussions, some of the interesting subjects
which can further reveal DM
properties: a) Many of the models predict anti-proton excesses at higher energies. Measurements of anti-proton with energy beyond the PAMELA range can reveal more details of DM b)  Almost all DM annihilation models predict certain level excesses of $\gamma$ ray. Further improved data on $\gamma$ ray can distinguish different models. c) Some of the models proposed, such as leptophlic models, predict very small cross sections for detection, but some other models with larger values. Direct detection of DM is certainly important to distinguish models. d) In some of the models there are light new particles. Negative results of direct detection of these particles can rule out some of these models. e) High energy collider search for DM or DM annihilation mediating particles or new light particles. LHC, ILC and CLIC can also play important role in distinguish different models.

We conclude that DM annihilation can provide a consistent explanation for the recently observed $e^\pm$ excesses in cosmic ray.
In order to cover the whole energy range of excesses observed, the DM mass must be as large as the highest energy observed showing excesses.
The FERMI-LAT and HESS data then require the DM to be around 1.5 to 2 TeV.
To produce large enough excesses with the annihilation mechanism, it requires modifications of the usual DM
properties because that the same DM annihilation process is also required to produce the relic DM density in the early universe.
With this constraint, a large boost factor in the range 100 to 1000 is needed to explain data. This boost factor can be provided by particle physics effects.
We have discussed two popular ones, the Sommerfeld and Breit-Wigner mechanisms. These two mechanisms have different consequences
which can be distinguished by future experimental observations. The Brei-Wigner mechanism requires that the annihilation is through s-channel and the mediating particle has a mass close to two times of the DM mass. While the Sommerfeld mechanism requires the existence of a light particle of mass less than a few GeV.
The PAMELA result of no anti-proton excesses further constrain how DM is annihilated. There are two classes of models, the leptophilic and the kinematically limited light particle decay models. The former requires the couplings cause the annihilation only have non-zero values for leptons, and the latter requires a new particle which the DM primarily annihilate into and this particle subsequently decays into leptons. Because this particle has a small mass which is kinematically forbidden to decay into proton or anti-proton and therefore explain the PAMELA null excess of anti-proton. All models modify what was called the usual WIMP DM. There are very different features for different types of models which can be tested. To further understand the properties of DM, more experimental observations are needed.

Acknowledgements: I thank Xiao-Jun Bi for useful discussions, and also Xiao-Jun Bi and and Qiang Yuan for collaborating on related subjects. This work was supported by NSC and NCTS.


\end{document}